\documentclass[8.5pt,twoside,twocolumn]{article}
\usepackage[super,sort&compress,comma]{natbib}
\usepackage{mhchem}
\usepackage{times,mathptmx}
\usepackage{sectsty}
\usepackage{balance}
\usepackage{graphicx} 
\usepackage{lastpage}
\usepackage[format=plain,justification=raggedright,singlelinecheck=false,font=small,labelfont=bf,labelsep=space]{caption}
\usepackage{fancyhdr}
\usepackage{color}
\definecolor{darkgreen}{rgb}{0,0.5,0}
\definecolor{blue}{rgb}{0,0,0.8}
\definecolor{lightblue}{rgb}{0.93,0.96,1}
\definecolor{darkblue}{rgb}{0.,0.,0.6}
\usepackage[colorlinks,linkcolor=darkgreen,citecolor=darkblue,urlcolor=blue,pdfborderstyle={/S/U/W 1}]{hyperref}

\begin{document}

\thispagestyle{plain}
\fancypagestyle{plain}{
\renewcommand{\headrulewidth}{1pt}}
\renewcommand{\thefootnote}{\fnsymbol{footnote}}
\renewcommand\footnoterule{\vspace*{1pt}%
\hrule width 3.4in height 0.4pt \vspace*{5pt}}
\setcounter{secnumdepth}{5}
\makeatletter
\def\subsubsection{\@startsection{subsubsection}{3}{10pt}{-1.25ex plus -1ex minus -.1ex}{0ex plus 0ex}{\normalsize\bf}}
\def\paragraph{\@startsection{paragraph}{4}{10pt}{-1.25ex plus -1ex minus -.1ex}{0ex plus 0ex}{\normalsize\textit}}
\renewcommand\@biblabel[1]{#1}
\renewcommand\@makefntext[1]%
{\noindent\makebox[0pt][r]{\@thefnmark\,}#1}
\makeatother
\renewcommand{\figurename}{\small{Fig.}~}
\sectionfont{\large}
\subsectionfont{\normalsize}

\fancyfoot{}
\fancyfoot[CO]{\vspace{-7.5pt}\hspace{0cm}\includegraphics{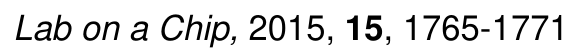}}
\fancyfoot[CE]{\vspace{-7.5pt}\hspace{0cm}\includegraphics{RF10}}
\fancyfoot[RO]{\footnotesize{\sffamily{ \hspace{2pt}\thepage}}}
\fancyfoot[LE]{\footnotesize{\sffamily{\thepage }}}
\fancyhead{}
\renewcommand{\headrulewidth}{1pt}
\renewcommand{\footrulewidth}{1pt}
\setlength{\arrayrulewidth}{1pt}
\setlength{\columnsep}{6.5mm}
\setlength\bibsep{1pt}

\twocolumn[
  \begin{@twocolumnfalse}
\noindent\LARGE{\textbf{Functional colloidal micro-sieves assembled and guided above a channel-free magnetic striped film$^\dag$}}
\vspace{0.6cm}

\noindent\large{\textbf{Fernando Martinez-Pedrero,\textit{$^{a}$} Arthur V. Straube,\textit{$^{b}$} Tom H. Johansen,\textit{$^{c,d}$} and
Pietro Tierno$^{\ast}$\textit{$^{a,e}$}}}\vspace{0.5cm}


\noindent {\small{Paper published in \textit{Lab on a Chip} \textbf{15}, 1765 (2015). DOI: \href{http://dx.doi.org/10.1039/c5lc00067j}{10.1039/c5lc00067j}}}
\vspace{0.6cm}

\noindent \textbf{{Abstract}}
\vspace{0.4cm}

\noindent \normalsize{
Colloidal inclusions in lab-on-a-chip devices can be used to perform
analytic operations in a non-invasive fashion.
We demonstrate here a novel
approach to realize fast and reversible micro-sieving operations
by manipulating
and transporting colloidal chains via mobile domain walls
in a magnetic structured
substrate.  We show that this technique allows to precisely
move and sieve non-magnetic
particles, to tweeze microscopic cargos or to mechanically
compress highly dense
colloidal monolayers.}
\vspace{0.5cm}
 \end{@twocolumnfalse}
  ]

\section{Introduction}
\footnotetext{\dag~Electronic Supplementary Information (ESI) --
Five movies (.AVI) illustrating the experiments and one file (.pdf) describing details of the theoretical model -- is freely available via
DOI: \href{http://dx.doi.org/10.1039/c5lc00067j}{10.1039/c5lc00067j}}
\footnotetext{\textit{$^{a}$~Departament de Estructura i Constituents de la Mat$\grave{e}$ria,
Universitat de Barcelona, Av. Diagonal 647, 08028 Barcelona, Spain. E-mail: ptierno@ub.edu}}
\footnotetext{\textit{$^{b}$~Institut f\"ur Physik, Humboldt-Universit\"at zu Berlin - Newtonstr. 15, D-12489 Berlin, Germany.}}
\footnotetext{\textit{$^{c}$~Department of Physics, The University of Oslo, P.O. Box 1048 Blindern, 0316 Oslo, Norway.}}
\footnotetext{\textit{$^{d}$~Institute for Superconducting and Electronic Materials, University of Wollongong, Northfields Avenue, Wollongong, NSW 2522, Australia.}}
\footnotetext{\textit{$^{e}$~Institut de Nanoci$\grave{e}$ncia i Nanotecnologia IN$^2$UB,Universitat de Barcelona, Barcelona, Spain.}}
Microfluidics, the art of handling nanoliter volume of reagents in
lithographically customized channel networks, has
direct applications in inorganic~\cite{Hassan2010} and analytical~\cite{Kim1995}
chemistry, biochemistry,~\cite{Kaigala2012} and life science.~\cite{Cab2005,Ohno2008} The ability
to perform complex operations within a micro-channel often
requires the use of ``active'' components, capable to control and
process small volumes of sample.~\cite{Xia1998,Isma2003}
The direct implementation of
mechanical units able to stir,~\cite{Yuen2003}
pump,~\cite{Laser2004} or sort~\cite{Kim2014} streams of fluids
in a single chip has been successfully demonstrated, although the
efficiency in device performance can be further optimized by combining
different strategies.~\cite{McD2003,Psaltis2006} In this context,
an alternative approach which
recently gained popularity relies on the use of micrometer scale colloidal
inclusions,~\cite{Terray2002} where single particles~\cite{Leach2006} or
small clusters of them~\cite{Bleil2006,Kavcic2009,Weddemann2010} can be
remotely actuated by an applied field, without direct mechanical contact.
Several basic functions can be performed in parallel or in a local fashion,
where the actuated particles are used as fluid stirrers, pumps, valves or
pistons.~\cite{Sawetzki2008}
Besides their addressability via external fields, another advantage
of implementing colloidal inclusions in lab-on-a-chip devices, is that the
particles can be used as individual drug delivery vectors once their surface
is chemically functionalized.~\cite{Matijevic2012} Static or low frequency oscillating  magnetic
fields are often preferred over other types of actuating forces due to their
non-invasive nature, and the fact
that biological systems remain practically unaffected.~\cite{Pamme2006}

Another important function in microfluidics system and, more in general, in
chemical engineering, is particle filtration. In lab-on-a-chip devices
this operation can be realized in different ways such as by incorporating
solid state membranes, or by creating pores via direct chemical etching or
photopolymerization, to cite a few methods.~\cite{Jong2006,Min2008,Lenshof2010,Didar2010,Didar2013} In most of the cases, however,
particle sieving has been achieved via static structures, characterized by
fixed and immobile reliefs, which could not be externally reconfigured at
will in order to stop or release the flow of matter.

Here we show an
alternative technique to move, sieve and trap colloidal cargos using
reconfigurable colloidal chains. These chains are formed and propelled
above a channel-free magnetic platform, allowing for an easy assembly
or disassembly by simply varying the applied magnetic field parameters.
\section{Materials and Methods}
\subsection{Colloidal particles}

As magnetic colloidal particles we use
monodisperse paramagnetic microspheres
from Invitrogen (Dynabeads M270),
composed of a highly cross-linked polystyrene matrix and
evenly doped with nanoscale superparamagnetic
grains (Fe$_2$O$_3$ and Fe$_3$O$_4$).
These particles are characterized by
a radius $a = 1.4~{\rm \mu m}$, a density
$\rho = 1.6~{\rm g~cm^{-3}}$ and a magnetic volume
susceptibility $\chi \approx 0.4$.~\cite{Helseth07}

As non-magnetic cargos, we use highly monodisperse micro-particles
based on silicon dioxide and having sizes ranging from $1$ to $5~{\rm \mu m}$,
which were purchased from Sigma Aldrich. These particles are diluted
in the deionized water and mixed at a proper concentration with the
paramagnetic colloidal particles before being deposited above the FGF.
\subsection{Magnetic film}

As a platform for the particle motion we use a
structured magnetic substrate, namely a ferrite garnet film (FGF) of
composition Y$_{2.5}$Bi$_{0.5}$Fe$_{5-q}$Ga$_q$O$_{12}$ ($q = 0.51$).
The FGF is grown by dipping liquid phase epitaxy on a gadolinium gallium
garnet substrate
from melt of the constituent rare earths containing bismuth, iron and
gallium, as well as PbO and B$_2$O$_3$.\cite{Tierno2009}
After successful growth, the FGF chip is characterized by
a regular lattice of parallel ferromagnetic stripe domains with
alternating perpendicular magnetization, and a spatial periodicity
of $\lambda = 2.5~{\rm \mu m}$ in zero applied field.
As shown in Fig.~\ref{Fig1},
separating these domains with opposite magnetization are Bloch
walls (BWs), i.e. narrow regions ($\sim 10 \, {\rm nm}$ width) which generate
strong gradients in the surface field. Moreover,
their positions can be manipulated by applying moderate
magnetic fields.
Before the experiments, the FGF film is coated with
a positive Photoresist AZ-1512 (Microchem, Newton, MA)
which is applied by using spin coating
(Spinner Ws-650Sz, Laurell) and photo-crosslinked
via UV irradiation (Mask Aligner MJB4, SUSS Microtec).
The complete procedure can be found in the
Supporting information of another article~\cite{Tierno2012}.
\subsection{Experimental procedure and system}

The magnetic colloidal particles
are dispersed in water and are electrostatically stabilized by the
negative charges acquired from the dissociation of the surface carboxylic
groups ($COO^{-}$). The original aqueous suspension of the
particles ($10~{\rm mg~ml^{-1}}$ or,
equivalently, $7\times10^9~{\rm beads~ml^{-1}}$)
is diluted with highly
deionized water ($18.2~{\rm M\Omega~cm}$, MilliQ system)
and few droplets are deposited above the FGF.

The motion of particles is recorded
by using a CCD camera (Balser Scout scA640-74fc) working at $75~{\rm fps}$.
The camera is mounted on top of a light microscope (Eclipse Ni, Nikon)
equipped with a $100\times$, $1.43~\rm{NA}$ objective and a TV adapter
having a magnification $0.45\times$. The optic system allows a total
field of view of $175\times109~\rm{\mu m^2}$. The positions of the
particles in the plane are obtained from the analysis of .AVI
movies recorded via a commercial software (STREAMPIX) and later
on analyzed using a MATLAB code based on the Crocker
and Grier software~\cite{Crocker1996}.

The precessing magnetic field is applied to the FGF by using
three custom-made solenoids arranged perpendicular to each
other and having the main axes along the $(x,y,z)$
directions, see Fig.~\ref{Fig1}. Two coils are connected to an AC power
amplifier (AKIYAMA AMP-1800) which is controlled by
a waveform generator (TGA1244, TTI) in order to
generate a rotating magnetic field in the $(x,z)$
plane. The third coil is connected to a DC power
supply (TTi El 302) to generate a static field along
the $y$ direction, $H_y$.
\section{Result and Discussion}
\subsection{Single particle motion}
\begin{figure}[t]
\includegraphics[width=\columnwidth,keepaspectratio]{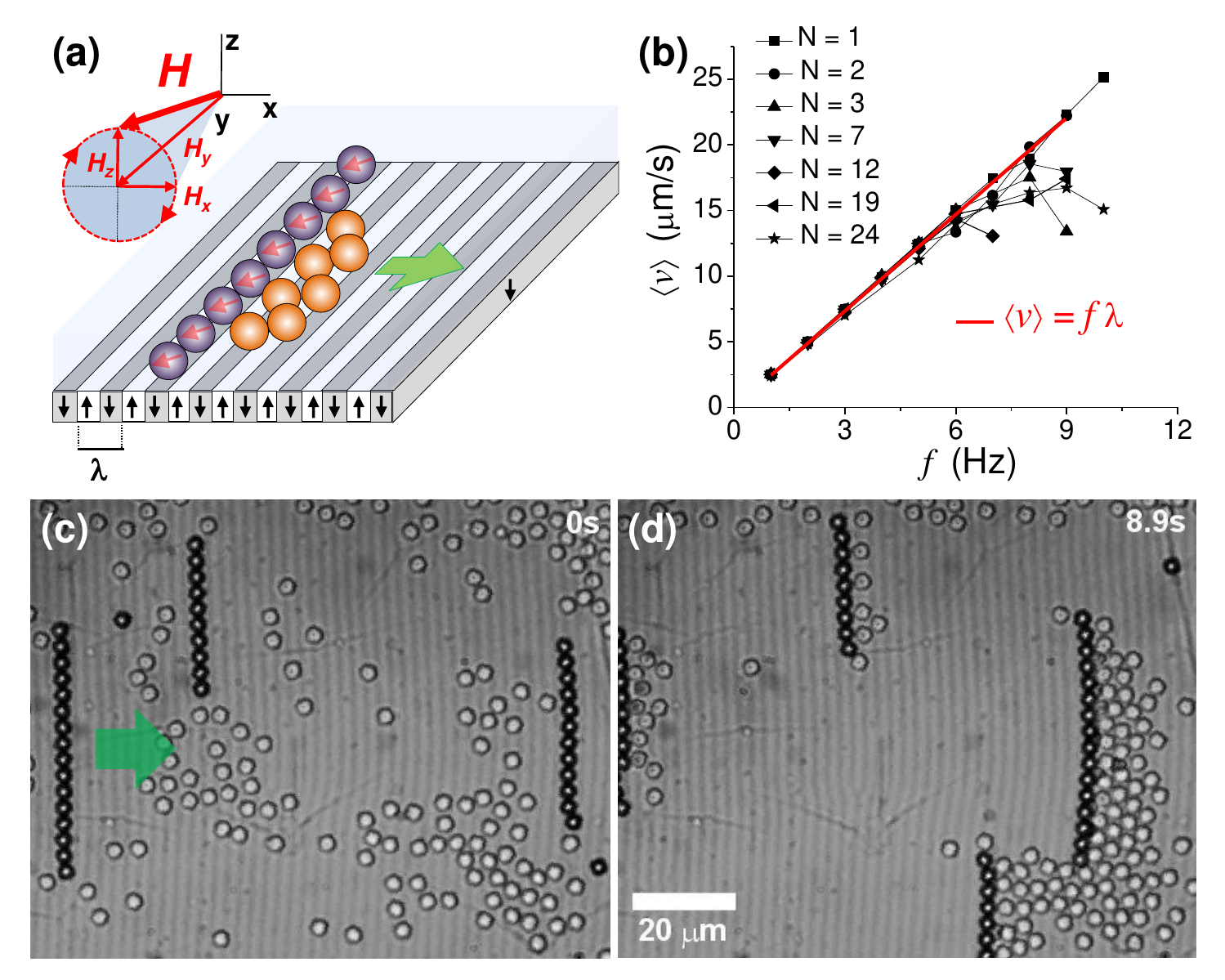}
\caption{(a) Schematic showing a chain of paramagnetic colloidal particles
transporting colloidal cargos above a FGF and under a precessing field.
(b) Average speed $\langle v \rangle$ of chains composed
by different number of particles ($N$) versus driving frequency $f$.
Red line denotes $\langle v \rangle= \lambda f$.
(c,d) Two images showing magnetic chains transporting
non-magnetic silica particles ($3~{\rm \mu m}$ size) at a speed of $7.5~{\rm \mu m~s^{-1}}$
($H_0 =1500~{\rm A~m^{-1}}$, $H_y=1500~{\rm A~m^{-1}}$, $f=3~{\rm Hz}$). See MovieS1
in the ESI.}
\label{Fig1}
\end{figure}

Our experimental system is illustrated schematically in
Fig.~\ref{Fig1}(a).
After being placed on the surface of the FGF chip,
the magnetic colloids are attracted by the BWs and become
confined in two dimensions due to balance between
magnetic attraction and steric repulsion.
The periodic arrangement of the BWs in the
FGF creates a one-dimensional
sinusoidal-like potential
along the $x$-direction (see Sec.~1 of the ESI).
Particle motion
is induced upon application of an external uniform magnetic
field rotating in the $(x,z)$ plane with frequency $f$ and amplitude $H_0$,
and with a field ellipticity denoted by $\beta \in[-1,1]$,
\begin{align}
\mathbf{H}^{\rm ext}=(H_x \cos(2\pi f t),0,-H_z \sin(2\pi f t)).
\end{align}
The two amplitudes $(H_x,H_z)$ are related to $(H_0,\beta)$ by,
\begin{align}
H_0=\sqrt{\frac{H_x^2 +H_z^2}{2}}\,, \quad \beta=\frac{H_x^2-H_z^2}{H_x^2+H_z^2}\,
\end{align}
such that $H_x=H_0\sqrt{1+\beta}$ and $H_z=H_0\sqrt{1-\beta}$.
Unless explicitly
stated, we used a circularly polarized magnetic field where $\beta=0$
and vary the frequencies $f\in [1, 20]~{\rm Hz}$ and the amplitude
$H_0 \in [500, 2500]~{\rm A~m^{-1}}$.
For these amplitudes,
the applied field $\mathbf{H}^{\rm ext}$ modulates the total surface
field of the FGF ($\mathbf{H}^{\rm sub}$) and
generates a traveling wave potential able to transport
the particles perpendicular to the stripe pattern
as shown in Fig.~S1 in the ESI.
For low frequencies,
the particles follow the magnetic potential,
and move at a constant
mean speed $\langle v \rangle = \lambda f$ above the FGF surface.
Increasing the driving frequency,
the particles reduce their average speed
due to the loss of synchronization
with the traveling potential.~\cite{Tierno20122,Straube2013}
\subsection{Magnetic chains and cargo transport}
To assemble the moving paramagnetic particles into
linear structures, we add to the rotating field a
static component of magnitude $H_y$, which causes the
field to perform a conical precession around the $y$-axis,
Fig.~\ref{Fig1}(a). The applied field now
reads as, $\mathbf{H}^{\rm ext}=(H_x \cos(2\pi f t),H_y,-H_z \sin(2\pi f t))$,
where $H_0$ continues indicating the amplitude of the rotating field
in the $(x,z)$ plane.
For amplitude $H_y > 0.7 H_0$ particles
located along the same BW experience attractive dipolar
interactions, and rapidly assemble into a one-dimensional chain moving
as a compact rod (see Sec. 2 of the ESI for details).
For low driving frequency, the average chain speed is
constant and proportional to $f$ via the relationship $\langle v \rangle=\lambda f$,
see Fig.~\ref{Fig1}(b). Under the same field parameters,
increasing the chain length decreases the maximum speed achievable due to
a corresponding increase in the friction coefficient
of the chain.

We used these magnetic chains to transport
colloidal cargos with sizes ranging from $2$ to to $5~{\rm \mu m}$.
Indeed, due to the comparable size of magnetic and non-magnetic
particles, individual paramagnetic particles
are unable to transport microscopic cargos unless the latter
are chemically bound to the particle surface.
In contrast, chains translating at a prescribed and well controlled speed
can be used to translate and accumulate
colloidal cargos randomly located above the FGF,
as shown in Fig.~\ref{Fig1}(c,d).
Once engaged into directed motion
by a propelling chain, the silica particles show negligible
thermal fluctuations with a small
diffusion coefficient in the transverse ($y$) direction,
$D_y\sim10^{-3}\,{\rm \mu m^2~s^{-1}}$ for particles having $3~{\rm \mu m}$ size.
This makes it difficult for them to escape by simply Brownian motion,
unless they are located exactly at the edge of the moving chain.
Although we demonstrate
this concept with colloidal particles,
other types of non-magnetic cargos
\begin{figure}[t]
\includegraphics[width=\columnwidth,keepaspectratio]{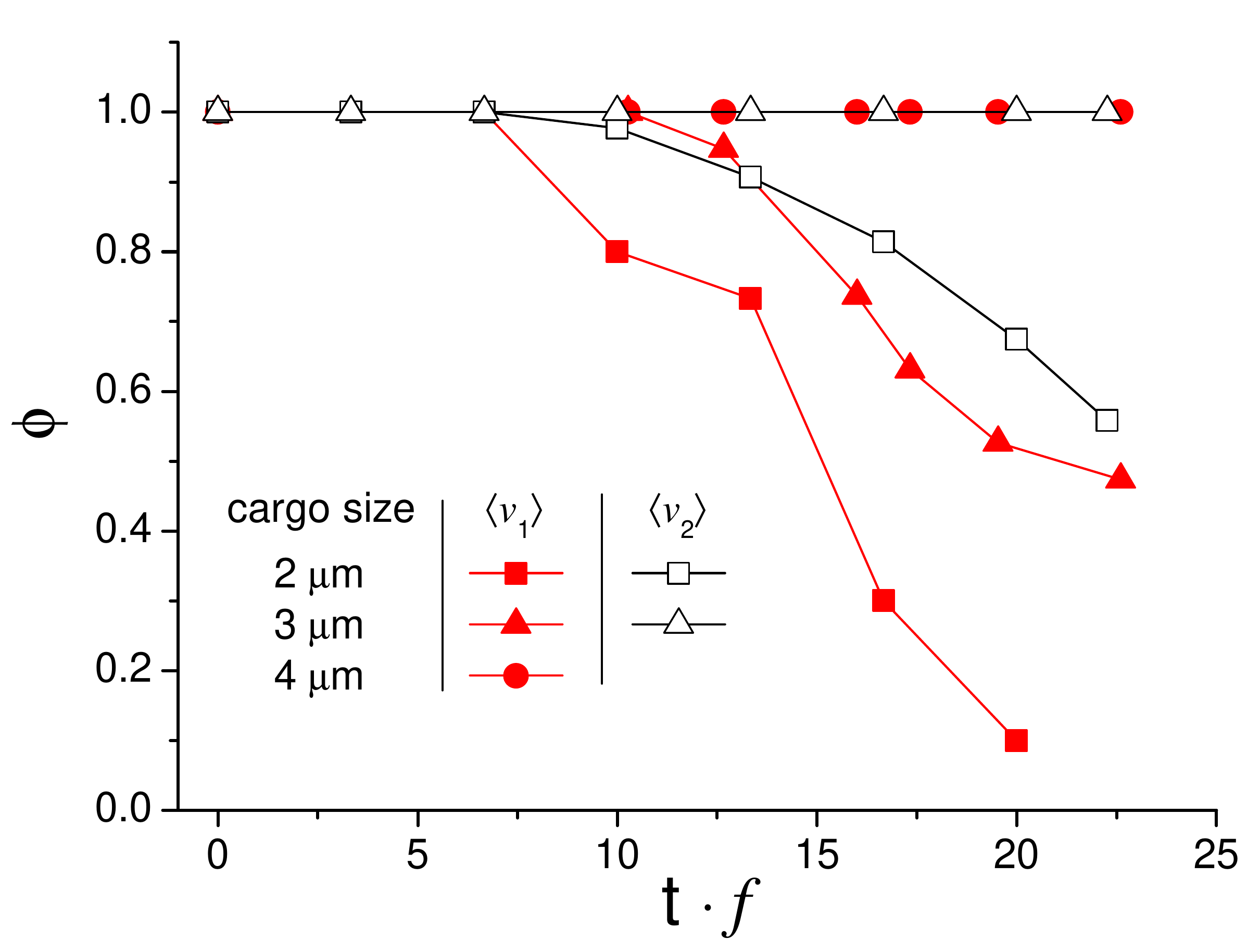}
\caption{Fraction of non-magnetic silica cargos
(``cargo'') $\phi$
transported by a moving chain of paramagnetic colloidal particles
versus normalized time $tf$
for three different sizes of the silica particles.
The driving frequencies used are $5~{\rm Hz}$
and $2\, Hz$
which correspond to average chain speeds $\langle v_1 \rangle=12.5~{\rm \mu m~s^{-1}}$
and $\langle v_2 \rangle=5~{\rm \mu m~s^{-1}}$, respectively.}
\label{Fig2}
\end{figure}
such as cells, liposomes or emulsion droplets can be equally well propelled.

In order to characterize the collection efficiency of
our magnetic chains in an open
system,
we perform a series of experiments by measuring
the fraction $\phi$ of non-magnetic cargos
which can be transported
over a given area $S \sim 2.8 \times 10^3~{\rm \mu m^2}$
by the chain at a
constant speed. The area in front of the chain decreases
with time as the chain propels.
As a consequence, as time proceeds
the colloidal cargos
start accumulating on one side of the moving chain.
The ideal case when all cargos are transported
by the moving chain corresponds to $\phi = 1$.
However, the cargos can escape from the moving chain
from the top part mainly for two reasons:
either due to diffusion in the perpendicular ($z$)
direction (which takes place only
for cargos having very small size),
or due to hydrodynamic lift
caused by an hydrodynamic perturbation
in the solvent generated by a fast translating barrier.
Thus in general the quantity $\phi \le 1$.
Fig.~\ref{Fig2} shows the evolution of $\phi$ versus time
for chains propelled by an applied field having amplitudes
$H_0= 1200~{\rm A~m^{-1}}$ and $H_y= 2300~{\rm A~m^{-1}}$
and different driving frequencies
which correspond to different speeds.
The time $t$ is normalized by the field frequency as $tf$,
in order to compare the different experiments
done at different frequencies.
The tendency of the smaller cargos
to cross the moving magnetic chain is
reflected by the decreases of $\phi$
starting from $t f\sim 10$ for cargos having
size $2~{\rm \mu m}$ and $3~{\rm \mu m}$.
Particles larger than $3~{\rm \mu m}$
can be transported along all
the area $S$ at any density up to their
close packing, $\phi=1$.
For a given size of cargos,
reducing the driving frequency increases
the collection efficiency since
the colloidal cargos have more time to redisperse within the
area reduced by the moving chain.

The limitation of our system to collect small particles
having size below $1~{\rm \mu m}$
can become an advantage
\begin{figure}[t]
\includegraphics[width=\columnwidth,keepaspectratio]{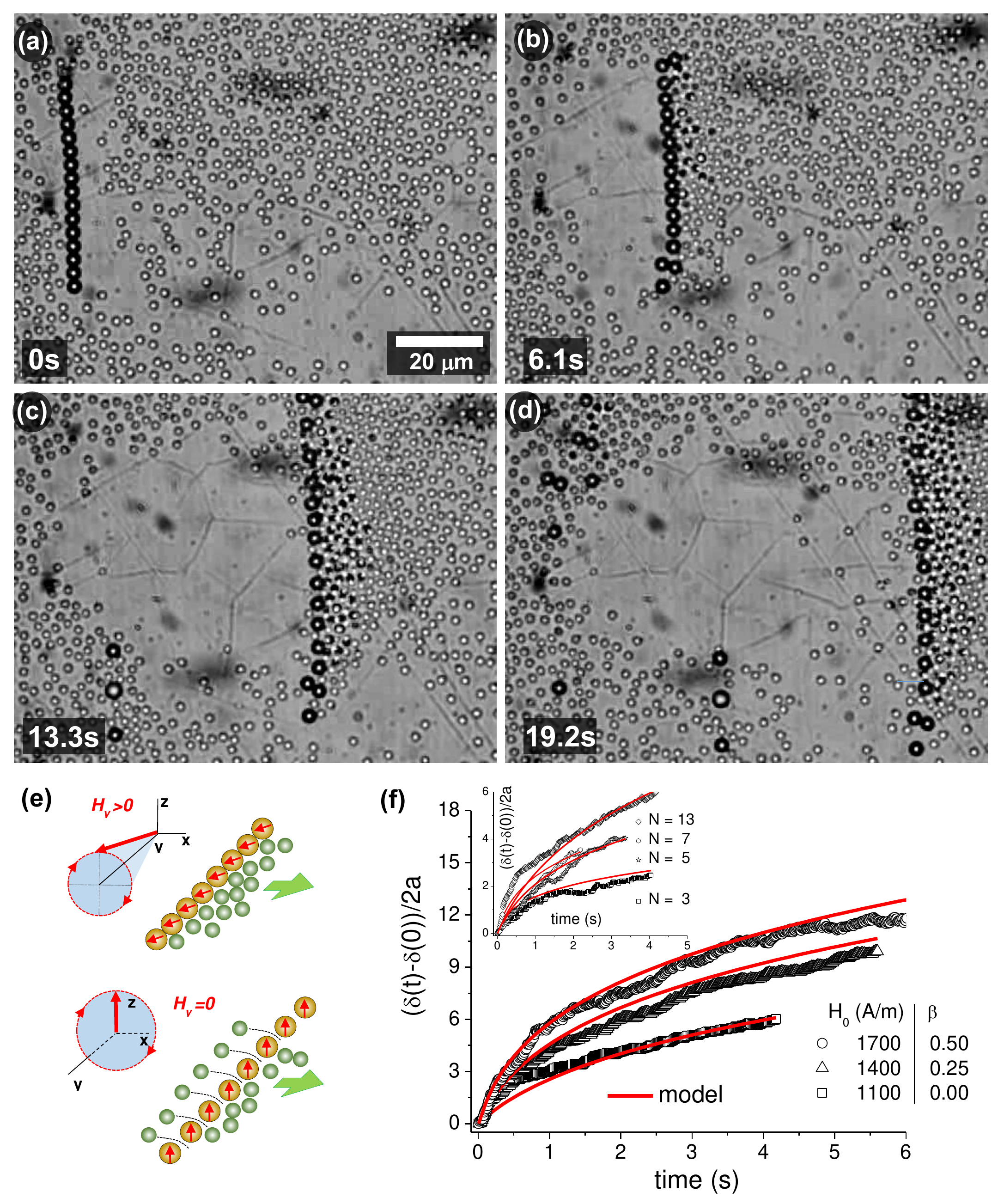}
\caption{(a-d) Images showing a chain of $17$ magnetic colloids
dragging and releasing $2~{\rm \mu m}$ silica
particles ($H_0=1500~{\rm A~m^{-1}}$, $H_y=2100~{\rm A~m^{-1}}$, $f=1.5~{\rm Hz}$).
Expansion of the chain was induced by switching off $H_y$
at $t=5.3~{\rm s}$, MovieS3 in the ESI. (e) Schematics showing
the process of chain opening. (f) Normalized end-to-end
distance $\delta(t)$ versus time after switching off
($t=0~{\rm s}$) the $H_y$ field for chain of $N=13$ particles at
different amplitudes $H_0$ of the rotating field.
Inset shows $\delta(t)$ for chains having different
lengths ($H_0=1100~{\rm A~m^{-1}}$).
Continuous red lines are fits following the theoretical model,
Eq.~(\ref{delta}).}
\label{Fig3}
\end{figure}
for other potential applications. For example,
it becomes relevant for sorting bi-disperse particle mixture
($1~{\rm \mu m}$ and $5~{\rm \mu m}$ size),
as shown in MoviesS2 in the ESI.
With this method,
we can capture and transport the larger
colloidal cargos, allowing for size
separation between fast and slowly
diffusing colloidal species deposited over the magnetic film.
\subsection{Micro-sieving}
Micro-sieving can be realized by tuning the dipolar interactions
which keep the particles in the chain. Fig.~\ref{Fig3}(a-d) shows
a sequence of images where an ensemble of silica particles ($2~{\rm \mu m}$ size)
are transported by a chain of $N=17$ magnetic particles (chain length $L=47.6~{\rm \mu m}$)
via a precessing field. After $t=5.3~{\rm s}$, the $y$-component of the field
is switched off, and the chain starts to open up and expand due
to repulsive interactions between the magnetic particles. After $\sim 7~{\rm s}$,
the average inter-particle distance, denoted as $\langle d \rangle$,
exceeds the size of the transported particles, which then start
flowing through the orifices driven by the pressure exerted
by the moving barrier. At $t=13~{\rm s}$, the expanded chain reaches a length of $\sim 80~{\rm \mu m}$
which corresponds to a mean inter-particle distance of $\langle d \rangle \sim 6.7~{\rm \mu m}$.
The flux of silica particles increases by increasing the
density of the compressed particles.
As shown in MovieS3, the chain can be easily recovered by switching on $H_y$
and the direction of motion reversed by changing the
sense of rotation of the field.

The schematics of Fig.~\ref{Fig3}(e) illustrate the chain
expansion mechanism. By switching off $H_y$,
the applied field now rotates in the plane perpendicular to the FGF
film, and the magnetic dipoles of the particles forming
the chain repel each other since they remain parallel
at all times. The transverse kinetics of the chain can be well
described by considering the balance between dipolar
and friction forces.
As derived in Sec.~3 of the ESI,
the average end-to-end distance of the chain $\delta(t)$
follows the law:
{
\begin{equation}
\delta(t)=2a(N-1)\left[ 1+  \frac{5\mu_0 \chi^2 H_0^2}{72 \eta F(z)(N-1)}\sum_{l=1}^{N-1}\frac{t}{l^4} \right]^{1/5}, \label{delta}
\end{equation}
}
where the permeability of water is $\mu_w \sim \mu_0 =4\pi \times 10^{-7}~{\rm H~m^{-1}}$
and $F(z)$ is a correction factor accounting for the proximity of the FGF
surface (see Eq.~(10) in the ESI). Since for long chains $\delta \sim (N-1)\langle d \rangle$,
$\langle d \rangle$ being the average inter-particle distance,
\begin{figure}[t]
\includegraphics[width=0.9\columnwidth,keepaspectratio]{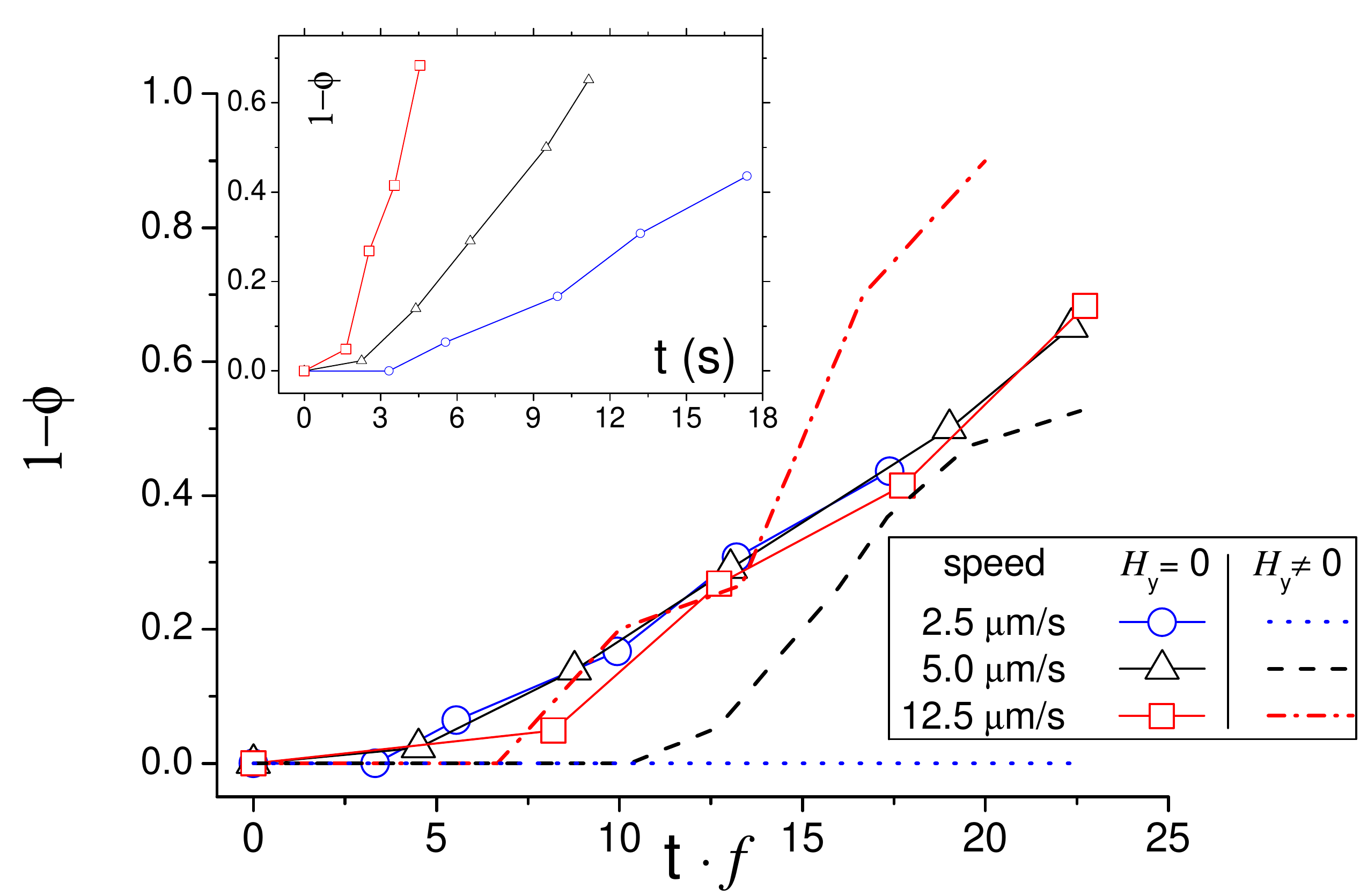}
\caption{Fraction of colloidal cargos ($1-\phi$)
which pass
a moving chain
versus normalized time $tf$
for three different
chain speeds.
Open symbols correspond
to sieving process ($H_y=0$)
while dashed and dotted lines
to non expanding chains ($H_y=2300~{\rm A~m^{-1}}$).
Inset shows the same data for
opened chains plotted as a function of the (non-rescaled) time.}
\label{Fig4}
\end{figure}
we can obtain an estimation of the average sizes of the pores.
Fig.~\ref{Fig3}(f) shows the good agreement between this theoretical
prediction and the experimental data for chains driven at
different amplitudes of the rotating field $H_0$.
The time $t=0~{\rm s}$ in the graph indicates the time when $H_y$
was switched off. The inset shows the results
obtained for chains composed of different number of particles.
Increasing the amplitude $H_0$ of the rotating field,
forces the propelling chain to expand faster, while decreasing
the number of particles in the chain increases the
expansion process for the same amplitude and
frequency of the applied field. In all cases,
at longer times, the average distance reaches
a plateau where the repulsion reduces significantly.
Since we can easily and independently change both $H_y$
and $H_0$, the chain expansion can be completely
controlled by the amplitude of the applied field,
allowing only the particles below the
pore size to pass through. Note that by setting $H_y =0.7 H_0$,
the expanding chain can be frozen at any
time as soon as it reaches the
configuration with a prescribed inter-particle distance.
\begin{figure*}[!t]
\begin{center}
\includegraphics[width=0.9\textwidth,keepaspectratio]{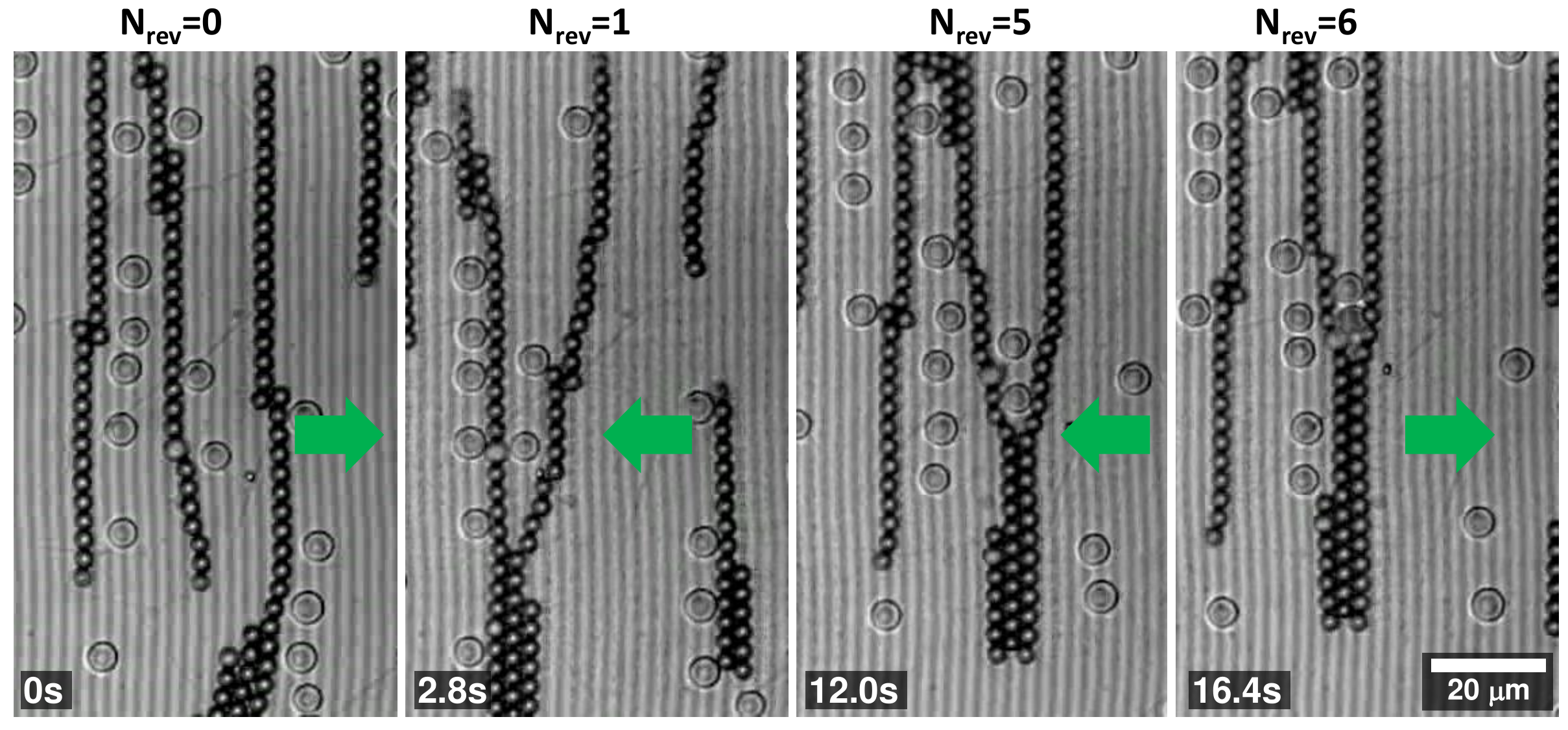}
\caption{Sequence of microscope images showing a pair of
chains entrapping three silica particles ($5~{\rm \mu m}$ size).
The external magnetic field has amplitudes
$H_0=1500~{\rm A~m^{-1}}$, $H_y = 1700~{\rm A~m^{-1}}$.
At the top of the images $N_{rev}$ indicates the
number of times the direction of motion
(green arrow) has been reversed, see also MovieS4 in the ESI.}
\label{Fig5}
\end{center}
\end{figure*}

To characterize the efficiency to sieve
non-magnetic particles
from the expanding pores,
we apply the same analysis as in Fig.~\ref{Fig2} but now reporting
the fraction $1-\phi$ of silica particles which
pass the open chain through the pores created
between the magnetic particles
when $H_y$ is switched off.
Fig.~\ref{Fig4} shows this fraction $1-\phi$ versus the normalized
time for three different
speeds when compressing $2\,\mu m$ size non-magnetic colloids.
Together with these data, the corresponding values for moving
compact chains in absence of expansion ($H_y = 2300~{\rm A~m^{-1}}$)
are shown as dashed and dotted lines.
The fact that the scattered points fall all on the same curve
rather than disperse, cf. the dotted and dashed lines,
shows that the
colloidal cargos pass the chain preferentially
through the pores rather than circumvent it from the top.
The possibility to accurately set the field strength and driving frequency
in order to tune the
size of the pores between the magnetic particles
allows us to speed-up, slow down or even
completely stop
the sieving process at any time,
a feature which is absent in static
membranes integrated
in microfluidics systems.

\subsection{Chain tweezing and compressing operations}
Besides sieving operations, a pair of
chains can be made attractive until
they clamp together performing particle ``tweezing''.
In particular, the interaction between two
moving particles along the direction
of motion perpendicular to the stripe
pattern can be tuned by varying the ellipticity $\beta$ of
the applied field. It was previously shown~\cite{Straube2014} that when
$\beta<-1/3$ the particles repel each other,
while magnetic attraction arises for $\beta>-1/3$.
We use this feature in Fig.~\ref{Fig5} to assemble
closely propelling chains into a colloidal ribbon for $\beta=0$.
The assembly process is induced after
moving first forward and later backward
the pair of chains, in such a way that the
sudden change in the direction of motion
causes the deformation of the chains.
During this process, one of the particles
forming the traveling chain, usually
located at one of the two ends, loses the
synchronization with the moving magnetic
potential. Adjacent particles are
forced to follow the retarded colloid,
and this delay in the propagation is continuously
transmitted along the chain. As a consequence,
the lateral distance between the two chains reduces,
while the latter approach each other. At a close distance,
attractive dipolar interactions assemble the chains
into a colloidal ribbon, merging via a
zip-like mechanism, and entrapping any particle
present between them, as shown in Fig.~\ref{Fig5}.

\begin{figure}[t]
\includegraphics[width=\columnwidth,keepaspectratio]{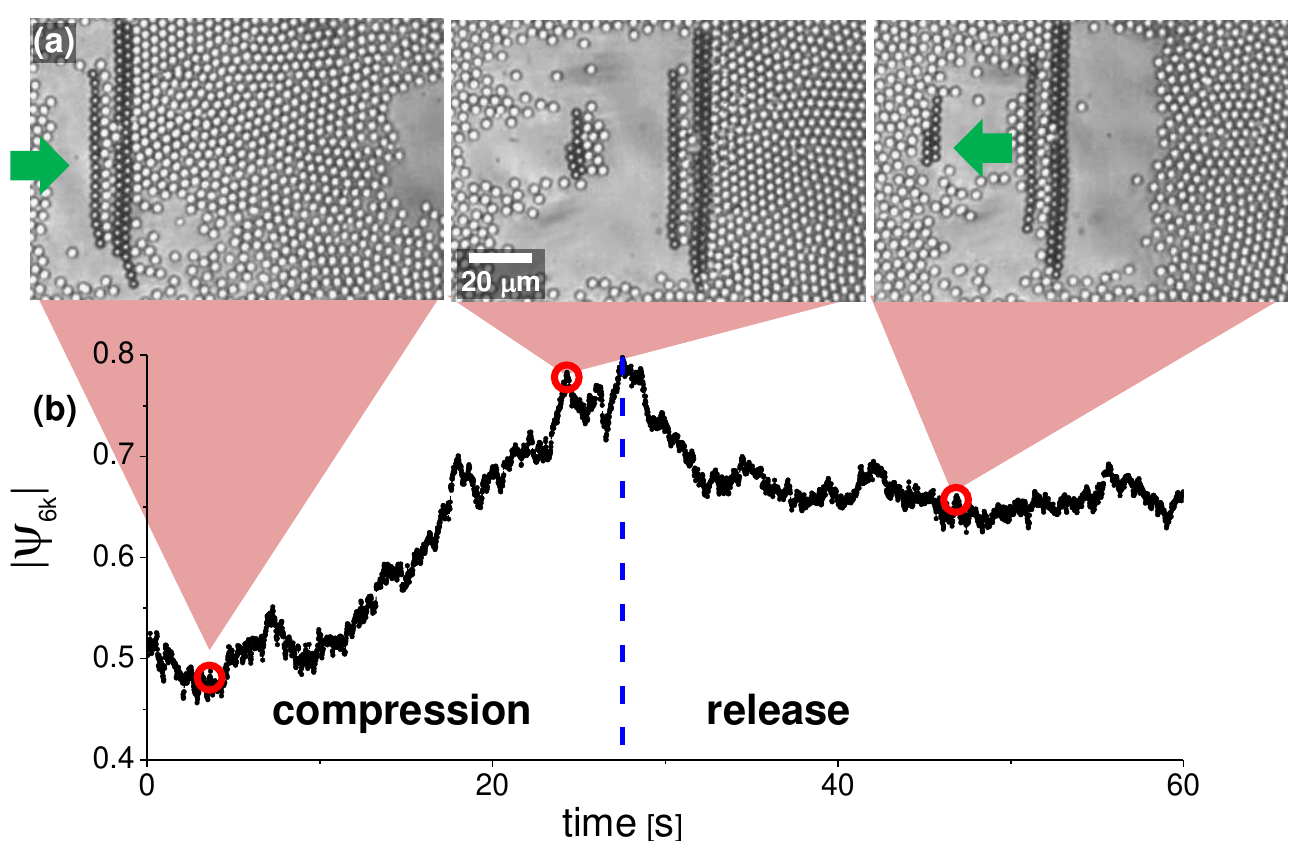}
\caption{(a) Sequence of images showing the compression of an
ensemble of silica particles ($3~{\rm \mu m}$ size) by a magnetic barrier
composed by two chains driven at a speed
of $\langle v \rangle = 2.5~{\rm \mu m~s^{-1}}$
($H_0 =1500~{\rm A~m^{-1}}$, $H_y=1500~{\rm A~m^{-1}}$, $f=1~{\rm Hz}$).
After $t=27.5~{\rm s}$ the direction of rotation of the field is inverted
($H_x$ to $-H_x$) and the chains move
backward, MovieS5. (b) Evolution of the orientational order
parameter of the colloidal monolayer.}
\label{Fig6}
\end{figure}

When a
pair of chains is stacked into a ribbon,
the latter is less prone to be deformed by the
colloidal cargo compared to the individual chains.
The magnetic ribbon can be thus used as a mobile barrier to compress
highly dense monolayer of non-magnetic particles, as shown
in Fig.~\ref{Fig6}(a) and MovieS5.
The system turns into an excellent model for studying
ordering in two dimensions,
where the surface pressure can be simply varied by moving the magnetic barrier along
the FGF surface. In order to characterize the solidification
(upon compression) and melting (upon release) of the monolayer,
we locate the particle positions via tracking routines, and measure the
$6$-fold bond-orientational order parameter, given
by $\psi_{6k}=\langle \exp[i 6 \theta_{kj} ]\rangle$.~\cite{Helseth2004}
Here $\theta_{kj}$ is the angle formed by a particle at location $k$ with its nearest neighbor $j$
with respect to a reference direction. For a perfect triangular lattice, $|\psi_{6k}|=1$.
As shown in Fig.~\ref{Fig6}(b), compression of the silica particles leads to the
formation of an ordered aggregate of particles in front of the chain ($|\psi_{6k}|\sim 0.8$)
even in absence of a hard wall confinement.
Inverting the direction of rotation of the field,
releases the monolayer and the particles start to slowly diffuse towards the empty area.
Melting of the lattice here is induced by thermal diffusion, and the orientational
order of the lattice starts to slowly decrease.
%

\section{Conclusions}
This article introduces a novel technique to
remotely manipulate and transport colloidal chains which can
be used to perform microscopic sieving and tweezing operations.
Permanently linked magnetic chains have been developed by various
groups in the past,~\cite{Philipyx1997,Biswal2003} and used
as micromechanical sensors,~\cite{Goubault2003}
fluid mixers~\cite{Furst1998} or stirrers~\cite{Ranzoni2010,Weddemann2011}
in lab-on-a-chip devices. In all these cases, however, the chains showed a
certain degree of flexibility~\cite{Petousis2007} which limited their use in analytic processes.
In contrast, in the present case, the magnetic substrate provides the ratchet
effect able to drive the particles and, at the same time, strongly pin the chain
position reducing its flexibility.
One limitation of our system is related to its inherent
two-dimensional nature, since the strong confinement
of the FGF restricts particle motion to a two-dimensional plane.
In order to increase the size of the chain, one could
use for example magnetic particles with rectangular shape\cite{Tav2013}.
An alternative strategy would be to stack FGFs on top of
each other in order to control the particle motion in parallel
planes as already proposed by other groups for different systems.\cite{Che2014}

Also, our approach allows to assemble (and disassemble)
magnetic chains via the use of a low-intensity external field, without need of any chemical
functionalization in order to link the particles. It should be also emphasized that the FGFs
are mechanically robust materials, inert to most chemicals, and transparent to visible light,
and can be integrated into any optical microscope system once they
are prepared with reduced dimensions (area $\sim 1~{\rm cm^2}$, thickness $\sim 50~{\rm \mu m}$).
The FGFs are also compatible with standard
soft-lithographic materials like PDMS as shown in the past.\cite{Tierno20072,Issle2008}
Finally, the presented technique can be potentially applied to
biological systems since, as previously reported,\cite{Dhar2007,Issle2008}
the FGF can be used to transport without damaging biological samples,
showing thus its promise in becoming a powerful tool for fluid based microscale technology.
%
\section{Acknowledgements}
F. M. P. and P. T. and acknowledge support
from the ERC starting grant ``DynaMO''
(No. 335040) and from the MEC via
programs No. RYC-2011-07605 and No.
FIS2011-15948-E. A.S. and P. T.
further acknowledge support from a
bilateral German-Spanish program of DAAD (project no. 57049473)
via the Bundesministerium f\"ur Bildung und Forschung (BMBF).

\footnotesize{
\bibliography{Biblio} 
\bibliographystyle{rsc} 
}
\end{document}